\documentclass[aps,prl,twocolumn,groupedaddress,showpacs]{revtex4}

\usepackage{graphicx}
\usepackage{dcolumn}
\usepackage{bm}
\usepackage{textcomp}

\begin{document}

\title{Quantum-Hall activation gaps in graphene}

\author{A.~J.~M.~Giesbers}
\email[]{J.Giesbers@science.ru.nl}
\affiliation{
High Field Magnet Laboratory, Institute for Molecules and Materials,
Radboud University Nijmegen, Toernooiveld 7, 6525 ED Nijmegen, The Netherlands
}

\author{U.~Zeitler}
\email[]{U.Zeitler@science.ru.nl}
\affiliation{
High Field Magnet Laboratory, Institute for Molecules and Materials,
Radboud University Nijmegen, Toernooiveld 7, 6525 ED Nijmegen, The Netherlands
}

\author{M.~I.~Katsnelson}
\affiliation{
Condensed Matter Theory, Institute for Molecules and Materials,
Radboud University Nijmegen, Toernooiveld 1, 6525 ED Nijmegen, The Netherlands
}

\author{L.~A.~Ponomarenko}
\affiliation{
Department of Physics, University of Manchester, M13 9PL, Manchester, UK
}

\author{T.~M.~G.~Mohiuddin}
\affiliation{
Department of Physics, University of Manchester, M13 9PL, Manchester, UK
}

\author{J.~C.~Maan}
\affiliation{
High Field Magnet Laboratory, Institute for Molecules and Materials,
Radboud University Nijmegen, Toernooiveld 7, 6525 ED Nijmegen, The Netherlands
}

\date{\today}

\begin{abstract}
We have measured the quantum-Hall activation gaps in graphene at
filling factors $\nu=2$ and $\nu=6$ for magnetic fields up to 32 T
and temperatures from 4 K to 300 K. The $\nu =6$ gap can be
described by thermal excitation to broadened Landau levels with a
width of 400 K. In contrast, the gap measured at $\nu=2$ is
strongly temperature and field dependent and approaches the
expected value for sharp Landau levels for fields $B > 20$ T and
temperatures $T > 100$ K. We explain this surprising behavior by a
narrowing of the lowest Landau level.
\end{abstract}


\maketitle

The quantum Hall effect (QHE) observed in two-dimensional electron
systems (2DESs) is one of the fundamental quantum phenomena in
solid state physics. Since its discovery in 1980 \cite{Klitzing}
it has been important for fundamental physics \cite{QHE-Book} and
application to quantum metrology \cite{metrology}. Recently a new
member joined the family of 2DESs: graphene, a single layer of
carbon atoms \cite{NovoselovScience,NovoselovPNAS,
Zhang_PRL,AndreNatMat,MishaMatTod}. Graphene displays a unique
charge carrier spectrum of chiral Dirac fermions
\cite{Semenoff,Haldane} and enriches the QHE with a half integer
QHE of massless relativistic particles observed in single-layer
graphene \cite{NovoselovNature,ZhangNature,Gusynin,CNeto} and a novel
type of integer QHE of massive chiral fermions in bilayers
\cite{bilayer,PRLFalko}. Moreover, the band structure of graphene
even allows the observation of the QHE up to room
temperature~\cite{RT_QHE}. Since localization in conventional
quantum Hall systems is already fully destroyed at moderate
temperatures, no QHE has been observed at temperatures above 30 K
until very recently. Therefore, understanding a room temperature
QHE in graphene goes far beyond our comprehension of the
traditional QHE.

In order to access this intriguing phenomenon in more detail we
report here systematic measurements of the inter Landau
level activation gap in graphene for magnetic fields up to 32 T.
We will show that the gap between the zeroth and the first Landau
level approaches the bare, unbroadened Landau-level separation for
high magnetic fields and we explain these findings by a much
narrower lowest Landau level compared to the other ones. In
contrast, for higher Landau levels, the measured activation gap
behaves as expected for equally broadened states.

\begin{figure}[t]  
  \begin{center}
 \includegraphics[width=0.9\linewidth]{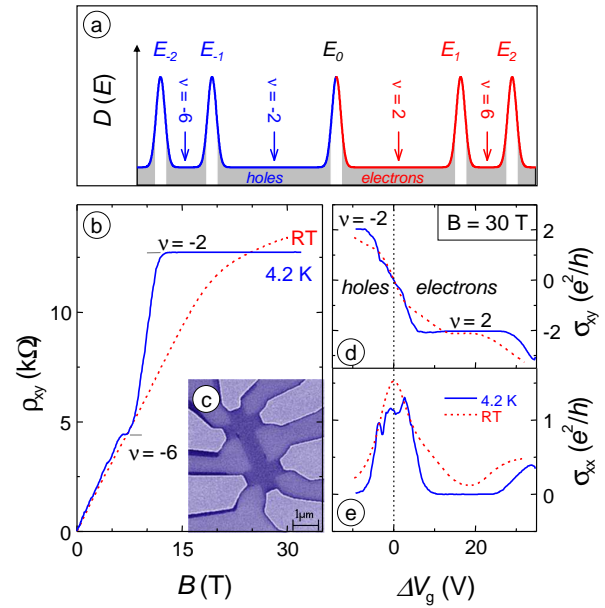}
  \end{center}
\caption{(Color online) (a) Schematic Landau-level structure in
graphene. The white areas are extended states in the center of the
Landau levels, with the localization radius of order of the sample
size, the gray areas represent localized states in between. The
arrows indicate the position of the chemical potential for the
corresponding filling factor.
\newline (b) Hall resistivity  $\rho_{xy}$ for holes in a single
layer graphene device as a function of the magnetic field. The
traces were recorded at 4.2 K (blue, solid) and RT (red, dashed)
at a fixed gate voltage corresponding to a carrier concentration
$n=-1.0 \times 10^{12} {\rm cm}^{-2}$.\newline (c) Scanning
electron micrograph of the graphene multiterminal device.\newline
Hall conductivity $\sigma_{xy}$ (d) and conductivity $\sigma_{xx}$
(e) at 4.2 K (blue, solid) and at RT (red, dashed) as a function
of the gate voltage at $B=30$~T.} \label{Figure1}
\end{figure}

The single-layer graphene samples  (Fig. \ref{Figure1}c) were
made by the micromechanical exfoliation of crystals of natural
graphite, followed by the selection of single-layer flakes using
optical microscopy and atomic force microscopy
\cite{NovoselovScience,NovoselovPNAS}. A large enough single-layer
flake is contacted by Au electrodes and patterned into a Hall bar
by e-beam lithography with subsequent reactive plasma etching. The
structures are deposited on a SIMOX-substrate with a 300 nm thick
SiO$_{2}$ layer on top of heavily doped Si. The Si is used as a backgate 
allowing to tune the carrier concentration $n$
to either holes ($n < 0$) or electrons ($n > 0$) with a
mobility $\mu=15000$~cm$^2$(Vs)$^{-1}$ at 4.2 K.
Due to the presence of surface impurities on the graphene 
sheet~\cite{SchedinNatMat}
the devices are generally strongly-hole doped with a
charge neutrality point situated at a positive back-gate voltage.
In order to restore a pristine undoped situation we anneal our samples 
at 390~K during several hours prior to any experiment; thereby removing 
most of the impurities and placing the charge neutrality point as close 
as possible to zero gate voltage~\cite{{AndreNatMat}}.

Electrons in graphene behave as chiral Dirac fermions with a
linear dispersion $E=c\hbar |k|$, where $c\approx 10^{6}~{\rm
ms^{-1}}$ is the electron velocity \cite{Haldane}. In a magnetic
field the energy spectrum splits up into non-equidistant Landau
levels (Fig.~\ref{Figure1}a) with energies given by
\cite{Gusynin,NovoselovNature,ZhangNature,CNeto}.
\begin{equation}\label{LLquantization}
E_{N} = {\rm sgn}(N)\; \sqrt{2\hbar c^2 eB \;|N|}.
\end{equation}
$N$ is a negative integer value for holes and positive for
electrons. The $N=0$ Landau level is shared equally between
both carrier types.

In Fig.~\ref{Figure1}b we have plotted the Hall
resistivity $\rho_{xy}$ as a function of the magnetic field for a
hole-doped device ($n=-1.0 \times 10^{12}$~cm$^{-2}$) at $T =
4.2$~K and at room temperature (RT). At 4.2 K pronounced plateaus,
accompanied by zero longitudinal resistivity, are visible in
$\rho_{xy}$ at values of $\rho_{xy} = -h/e^{2}\nu$, with
$\nu=-2$ and $\nu=-6$. At $B\approx 21 \;{\rm T}$ ($\nu=-2$) the
Fermi energy is situated on the localized states between the
zeroth ($N=0$) and first ($N=-1$) Landau level of the holes (see
Fig.~\ref{Figure1}a). All levels below the Fermi energy are now
completely filled and $\rho_{xy}$ is quantized to
$h/2e^{2}$. When sweeping the magnetic field downwards more Landau
levels become populated and $\nu=-6$ is reached at $B\approx 7$~T.
The Fermi energy now lies between the first ($N=-1$)
and the second ($N=-2$) Landau level and $\rho_{xy}$ is
quantized at $h/6e^{2}$.

The quantization of the Hall resistance, especially at room
temperature, can be seen most clearly by varying the carrier
concentration at the highest possible magnetic field. This is
shown in Fig.~\ref{Figure1}d and Fig.~\ref{Figure1}e for $B=30$~T,
where we plot the Hall conductivity $\sigma_{xy}$ and the conductivity
$\sigma_{xx}$ as a function of the applied gate voltage
at 4.2 K and RT .
The conductivity tensor $\bm{\sigma}$ was calculated by inverting
the experimentally measured resistivity tensor $\bm{\rho}$.
The corresponding quantum Hall plateaus for holes ($\nu=-2$) and electrons ($\nu=2$),
accompanied by minima in  $\sigma_{xx},$
are quantized to $\sigma_{xy}=\pm2e^{2}/h$ and remain visible up to
RT.

It is interesting to note that the low-temperature data shown in
Fig.~\ref{Figure1}e demonstrate a splitting of the conductivity
maximum around the charge neutrality point which
disappears at higher temperatures.
As reported previously~\cite{Zhang_PRL, Abanin_PRL}, this observation
may be explained as a Zeeman splitting of the order of a few Kelvin~\cite{Zhang_PRL}
or the presence of counter propagating edge channels dominating the resistivity
$\rho_{xx}$~\cite{Abanin_PRL}.
Additionally, a spontaneous spin and/or valley
polarization due to $SU(4)$ quantum Hall ferromagnetism was suggested \cite{nomura}, possibly
explaining the disappearance of the splitting at higher temperatures with a crossing
of the Curie temperature.
Whether such a polarization survives up to the high
temperatures with a value of the splitting smaller than
the thermal energy is still subject of scientific debate and goes beyond the scope
of this work.
For completely filled Landau levels, however, as we consider in the rest of this paper,
no spontaneous polarization takes place and no interaction induced splitting is expected.

\begin{figure}[t]  
 \begin{center}
 \includegraphics[width=0.9\linewidth]{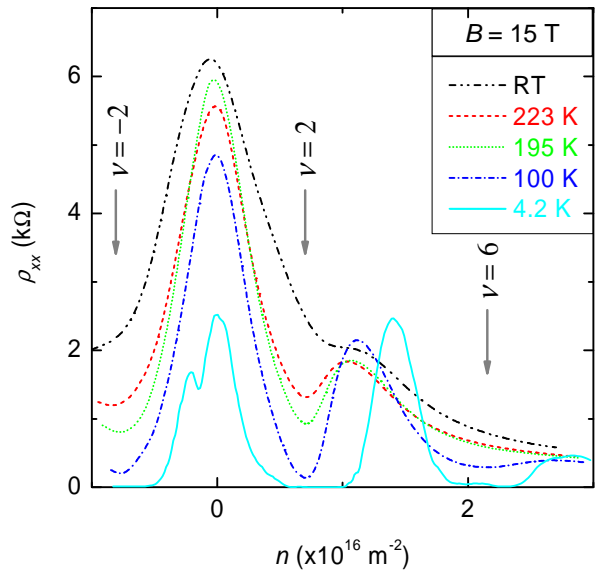}
  \end{center}
\caption{(Color online) Resistivity $\rho_{xx}$ as a function of carrier
concentration at $B=15$~T for different temperatures.
The arrows mark the  $\nu=\pm 2$ and the $\nu=6$ minima.}
  \label{Figure2}
\end{figure}

To determine the energy gaps we have measured $\rho_{xx}$
as a function of the carrier concentration at a
constant magnetic field (5 to 30 T in steps of 5 T) and
at different temperatures between 4.2 K and RT. Typical results for
$B=15$~T are shown in Fig.~\ref{Figure2}. 
At this intermediate field all the minima already display a clearly visible temperature dependence. At the highest field (30 T) the $\nu=2$ minimum 
remains close to zero for all temperatures (see Fig.~1e).
The value of the minima in the longitudinal resistivity starts to deviate
from zero with increasing temperature and the quantum Hall
plateaus become less pronounced. Clearly, the $\nu=\pm 2$ minimum
is much more robust at higher temperatures than the $\nu=6$
minimum, which is also the case for the related quantum Hall
plateaus.

\begin{figure}[t]  
  \begin{center}
  \includegraphics[width=0.9\linewidth,angle=0]{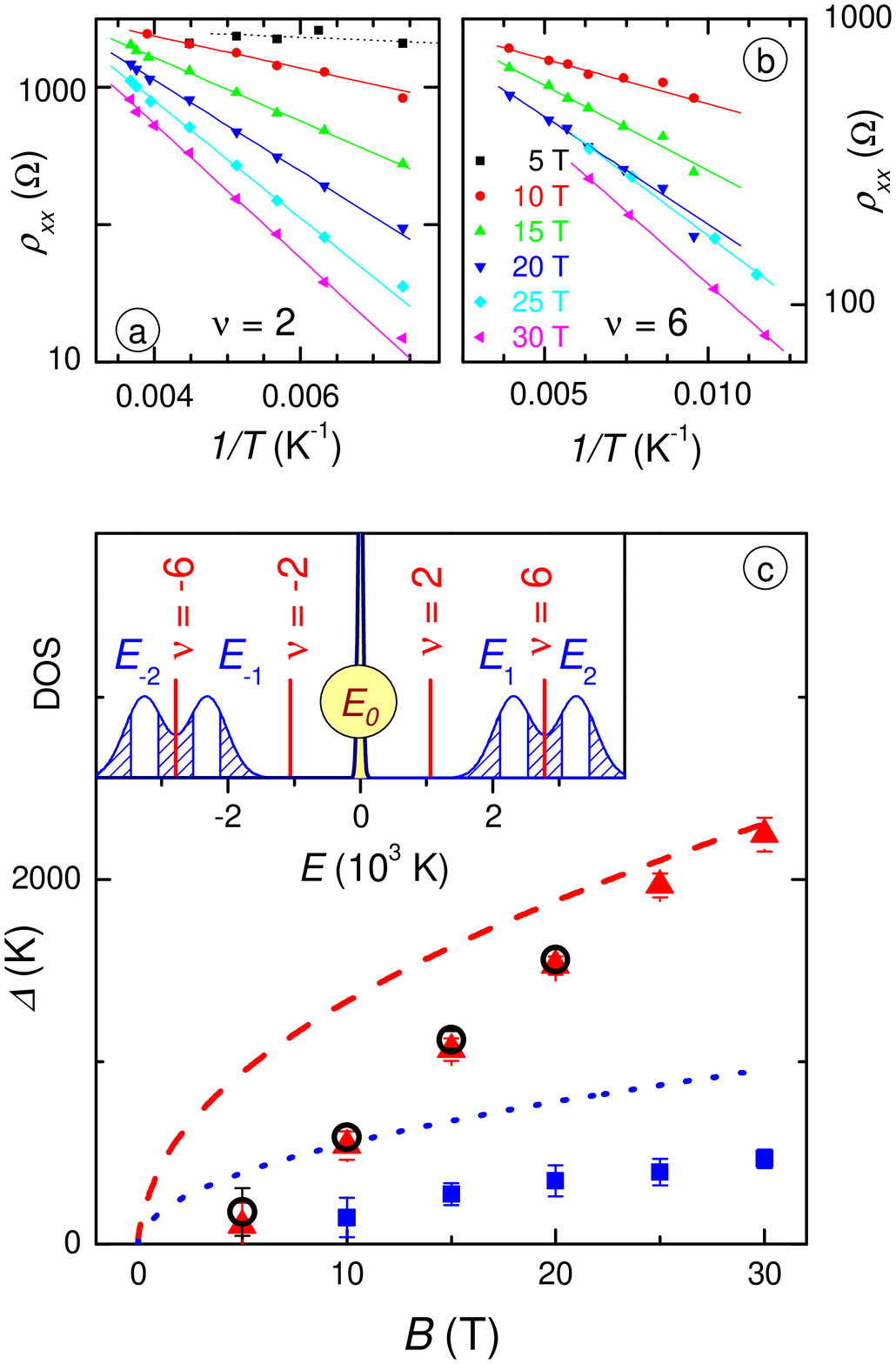}
  \end{center}
\caption{(Color online) Arrhenius plots of $\rho_{xx}$ in the high
temperature range for the gaps at $\nu=2$ (a) and at $\nu=6$ (b)
for different magnetic fields. \newline (c) Energy gaps
$2\Delta_{a}$ between two Landau levels as a function of magnetic
field for $\nu=+2$ (full red triangles) $\nu=-2$ (open black
circles) and $\nu=6$ (full blue squares) as deduced from the
Arrhenius plots. The dashed (red) and dotted (blue) lines are the
theoretically expected energy gaps for sharp Landau levels. The
insert shows schematically the density of states for a sharp
zeroth Landau level and broadened higher Landau levels for
electrons and holes at 30 T. The form and width of the higher
Landau levels were extracted from experimental data. Extended
states are represented by the white areas, localized states by the
dashed areas. }
  \label{Figure3}
\end{figure}

The value of the resistance minima at $\nu=2$ and $\nu=6$ as a
function of the inverse temperature for different magnetic fields
are shown in Fig.~\ref{Figure3}a and Fig.~\ref{Figure3}b
respectively~\cite{remark_nu6}. 
From the slope in these Arrhenius plots,
\begin{equation}\label{gap}
  \rho_{xx} \propto \exp{(-\Delta_{a}/kT)},
\end{equation}
we deduce the activation gap $\Delta_{a}$ 
All data for $T \geq 100 \; {\rm K}$ can be reasonably fitted with a single
Arrhenius exponent. 
For $T<100$~K, in particular for $\nu=\pm2$, the Arrhenius
plots flatten off significantly, an effect normally attributed 
to variable-range-hopping~\cite{VRH1,VRH2}.

In Fig.~\ref{Figure3}c we show the results of the measured gaps
and compare them to the bare Landau-level separation as given by
Eq.(\ref{LLquantization}). 
Gaps for $\nu = -2$ were 
obtained using a similar Arrhenius analysis as in Fig.~3a, 
yielding very comparable results as for the gaps at $\nu=+2$.
However, due to a leakage current through the gate insulator
for too negative gate voltages, $\nu =-2$ moves beyond experimental
reach for $B>20$~T.

The experimental values for $\nu=6$
show a constant offset of $\sim400 \; {\rm K}$ from the ideal
curve, which can straightforwardly be explained by a corresponding
finite Landau-level width~\cite{Ando}.

In contrast, the $\nu=\pm2$ gap behaves strikingly different. At
low magnetic fields a lower value than expected for an ideal
system is measured, but for high magnetic fields the measured gap
approaches the bare Landau-level separation. Since the results at
$\nu=6$ show that the $N=1$ Landau level behaves as expected, this
peculiar behavior at $\nu=\pm2$ can only be explained by the
unique nature of the $N=0$ Landau level shared equally between
electrons and holes of opposite chirality.

In order understand this behavior in more detail we 
have to consider that the measured activation gap 
is determined by the distance from the chemical potential to the conductivity edges of the two
adjacent Landau levels. When these levels have considerably different 
mobilties or widths (a case normally not encountered in traditional QHE samples),  
this gap will be dominated by the distance of the Fermi energy
to the nearest Landau level with the highest mobility minus half its width.
Therefore, in our case the conductivity around $\nu=2$ is dominated by thermal excitation to the $N=0$ Landau level since the peak conductivity measured in density sweeps is considerably larger for the $N=0$ Landau level than for $N=1$ (see Fig.~\ref{Figure1}e). 
In particular, when this level is narrow enough,
the high field excitation gap then just reflects the bare Landau level separation. 

Such a narrowing of the lowest Landau level is also supported theoretically by the following arguments: A major source of disorder in graphene can be tracked down
to its corrugated (rippled) surface structure~\cite{morozov,jannik,stolyarova,SSC} leading
to random fluctuations of the perpendicular magnetic field and,
as a consequence, a considerable broadening of higher  Landau levels. 
The zero-energy level, however, is exceptional in this sense. It is
topologically protected by the so-called Atiyah-Singer index
theorem, such that the number of states with zero energy is only
determined by the total magnetic flux through the system and does
not depend on whether this field is uniform or
not~\cite{NovoselovNature,MishaMatTod}. Therefore, the
fluctuations of the vector potential caused by ripples
are not able to broaden this zero-energy state. The special nature
of this state is most pronounced in high magnetic fields where the
lowest Landau level is well separated form the neighboring levels;
for lower fields Landau level mixing can broaden the lowest Landau
level by means of inter Landau-level scattering.

For a more quantitative analysis we estimate the density of states in the higher Landau levels $N=1$ and $N=2$ from the measured resistivity as a function of concentration. Identifying localized states with a zero low-temperature resistivity $\rho_{xx}$ and extended states with non-zero $\rho_{xx}$ we find that the number of localized states between the first and second Landau level is about 2.5 times the number of extended states in one of the levels. Knowing the width of the extended states $\Gamma = 400$~K, as deduced above, the form of the first and second Landau level is fully determined and quantitatively sketched in the inset of Fig.~\ref{Figure3}c.

Using this density of states for the $N=1$ level and a sharp $N=0$ level allows us to calculate the field and temperature dependence of the chemical potential $\epsilon_F$ at $\nu=\pm2$ using standard Fermi statistics.
Above 100 K and in high magnetic fields $\epsilon_F$ is found to be near the middle of the gap.
The conductivity at $\nu=\pm2$, is then dominated by thermal excitation to the $N=0$ Landau level (with the highest mobility) which amount to half the Landau level distance,
a value we indeed measure experimentally. 
These findings also agree with our Arrhenius plots in Fig.~\ref{Figure2} with a well defined single slope for $T>100$~K.

Interestingly, our proposed scenario of an asymmetric density of states around
$\nu=2$ also implies a reduction of the Arrhenius slope for $T<100$~K. It forces the Fermi energy $\epsilon_F$ to move from a mid-gap position closer towards the lowest Landau level. This effect indeed reduces the low-temperature activation energies and is independent from other effects such as variable-range hopping which can play a similarly important role for the reduction of the measured gaps.

Finally, we note that the measured gaps are extremely sensitive to
background doping from surface impurities. 
Exposing the sample
to air, hereby absorbing surface impurities, induces 
extra doping and considerably reduces the measured gaps. In particular,
a narrowing of the lowest Landau level can no longer be observed. 
This statement is also confirmed by the fact that the strong temperature dependence
of the $\nu = 2$ gap disappears for a system with surface impurities.

In conclusion, we have measured the excitation gaps at $\nu=\pm2$
and $\nu=6$ in graphene. We have shown that the results for higher
Landau levels can be described quantitatively by thermal
activation to broadened Landau levels with a large Landau-level
width $\Gamma \approx 400$~K for the sample investigated. The
lowest Landau level, however, becomes very sharp with increasing
magnetic field, and the gap approaches the bare Landau-level
splitting.


Part of this work has been supported by EuroMagNET
under EU contract RII3-CT-2004-506239 and by the Stichting Fundamenteel
Onderzoek der Materie (FOM) with financial support from the
Nederlandse Organisatie voor Wetenschappelijk Onderzoek (NWO).



\begin{thebibliography}{99}


\bibitem{Klitzing}
K.~v.Klitzing, G.~Dorda, and M.~Pepper,
Phys.~Rev.~Lett.~\textbf{45}, 494 (1980).


\bibitem{QHE-Book}
See, e.g., D.~Yoshioka,
{\sl The Quantum Hall Effect} (Springer, Berlin, 2002).


\bibitem{metrology}
B.~Jeckelmann and B.~Jeanneret,
Rep.~Prog.~Phys.~\textbf{64}, 1603 (2001).


\bibitem{NovoselovScience}
K.~S.~Novoselov \emph{et al.},
Science \textbf{306}, 666 (2004).


\bibitem{NovoselovPNAS}
K.~S.~Novoselov \emph{et al.},
Proc.~Natl.~Acad.~Sci.~USA \textbf{102}, 10451 (2005).


\bibitem{Zhang_PRL}
Y.~Zhang \emph{et al.},
Phys.~Rev.~Lett.~\textbf{96}, 136806 (2006).


\bibitem{AndreNatMat}
A.~K.~Geim and K.~S.~Novoselov,
Nature Materials \textbf{6}, 183 (2007).


\bibitem{MishaMatTod} M.~I.~Katsnelson,
Mater. Today {\bf 10}(1-2), 20 (2007).


\bibitem{Semenoff}
G.~W.~Semenoff,
Phys.~Rev.~Lett.~\textbf{53}, 2449 (1984).


\bibitem{Haldane}
F.~D.~M.~Haldane,
Phys.~Rev.~Lett.~\textbf{61}, 2015 (1988).


\bibitem{NovoselovNature}
K.~S.~Novoselov \emph{et al.},
Nature \textbf{438}, 197 (2005).


\bibitem{ZhangNature}
Y.~Zhang, Y.~Tan, H.~L.~Stormer, and P.~Kim,
Nature \textbf{438}, 201 (2005).


\bibitem{Gusynin}
V.~P.~Gusynin and S.~G.~Sharapov,
Phys.~Rev.~Lett.~\textbf{95}, 146801 (2005).

\bibitem{CNeto}
N.~M.~R.~Peres, F.~Guinea, and A.~H.~Castro~Neto,
Phys.~Rev.~B~\textbf{73}, 125411 (2006)

\bibitem{bilayer}
K.~S.~Novoselov \emph{et al.}
Nature Phys. \textbf{2}, 177 (2006).


\bibitem{PRLFalko}
E.~McCann and V.I.~Fal'ko,
Phys.~Rev.~Lett.~\textbf{96}, 086805 (2006).


\bibitem{RT_QHE}
K.~S.~Novoselov \emph{et al.},
Science \textbf{315}, 1379 (2007).

\bibitem{SchedinNatMat}
F.~Schedin \emph{et al.},
Nature Materials {\bf 6}, 652 (2007). 


\bibitem{Abanin_PRL}
D.~A.~Abanin \emph{et al.},
Phys.~Rev.~Lett.~\textbf{98}, 196806 (2007).


\bibitem{nomura}
K.~Nomura and A.~H.~MacDonald,
Phys.~Rev.~Lett.~{\bf 96}, 256602 (2006).

\bibitem{remark_nu6}
The gaps at $\nu=6$ were measured in two different cooldowns, as a
result the 25 and 30 T lines in Fig.~\ref{Figure3}b are slightly
shifted upwards. In order to confirm that the measured gaps at
high fields were still in agreement with the rest of the
measurements we compared the gap at 15 and 20 T for both cooldowns
and found that they showed an almost identical value.

\bibitem{VRH1}
Y.~Ono {\sl et al.}, 
J. Phys. Soc. Japan {\bf 51}, 237 (1982).

\bibitem{VRH2}
G.~Ebert {\sl et al.},
Solid State Commun. {\bf 45}, 625 (1983).

\bibitem{Ando}
T.~Ando, A.~B.~Fowler, and F.~Stern,
Rev.~Mod.~Phys.~\textbf{54}, 437 (1982).





\bibitem{morozov} 
S.~V. Morozov {\sl et al.},
Phys. Rev. Lett. {\bf 97}, 016801 (2006).


\bibitem{jannik}
J.~C.~Meyer {\sl et al.},
Nature {\bf 446}, 60  (2007); Solid State Commun. {\bf 143}, 101
(2007).

\bibitem{stolyarova}
E. Stolyarova {\sl et al.}, 
Proc.~Natl.~Acad.~Sci.~USA \textbf{104}, 9209 (2007).

\bibitem{SSC}
M. I. Katsnelson and K. S. Novoselov, 
Solid State Commun. {\bf 143}, 3 (2007).



\end{thebibliography}
\end{document}